\documentclass[aps,pra,reprint,superscriptaddress]{revtex4-2}

\usepackage{graphicx}
\usepackage{color}
\usepackage{amsmath,amssymb}
\usepackage{bm}
\usepackage{upgreek}
\usepackage{xspace}
\usepackage{bbold}
\usepackage[colorlinks,urlcolor=blue,citecolor=blue,linkcolor=blue]{hyperref}
\catcode`_=12
\begingroup\lccode`~=`_\lowercase{\endgroup\let~\sb}
\renewcommand{\selectlanguage}{}
\usepackage[colorlinks,urlcolor=blue,citecolor=blue,linkcolor=blue]{hyperref}
\newcommand{\uvect}[1]{\hat{\boldsymbol{#1}}\xspace}

\begin{document}

\title{Quantum control of nuclear spin qubits in a rapidly rotating diamond}

\author{A.~A.~Wood}
\email{alexander.wood@unimelb.edu.au}
\affiliation{School of Physics, University of Melbourne, Victoria 3010, Australia}
\author{R. M. Goldblatt}
\affiliation{School of Physics, University of Melbourne, Victoria 3010, Australia}
\author{R. E. Scholten}
\affiliation{School of Physics, University of Melbourne, Victoria 3010, Australia}
\author{A. M. Martin}
\affiliation{School of Physics, University of Melbourne, Victoria 3010, Australia}

\date{\today}
\begin{abstract}
Nuclear spins in certain solids couple weakly to their environment, making them attractive candidates for quantum information processing and inertial sensing. When coupled to the spin of an optically-active electron, nuclear spins can be rapidly polarized, controlled and read via lasers and radiofrequency fields. Possessing coherence times of several milliseconds at room temperature, nuclear spins hosted by a nitrogen-vacancy center in diamond are thus intriguing systems to observe how classical physical rotation at quantum timescales affects a quantum system. Unlocking this potential is hampered by precise and inflexible constraints on magnetic field strength and alignment in order to optically induce nuclear polarization, which restricts the scope for further study and applications. In this work, we demonstrate optical nuclear spin polarization and rapid quantum control of nuclear spins in a diamond physically rotating at $1\,$kHz, faster than the nuclear spin coherence time. Free from the need to maintain strict field alignment, we are able to measure and control nuclear spins in hitherto inaccessible regimes, such as in the presence of a large, time-varying magnetic field that makes an angle of more than $100^\circ$ to the nitrogen-lattice vacancy axis. The field induces spin mixing between the electron and nuclear states of the qubits, decoupling them from oscillating rf fields. We are able to demonstrate that coherent spin state control is possible at any point of the rotation, and even for up to six rotation periods. We combine continuous dynamical decoupling with quantum feedforward control to eliminate decoherence induced by imperfect mechanical rotation. Our work liberates a previously inaccessible degree of freedom of the NV nuclear spin, unlocking new approaches to quantum control and rotation sensing.  

\end{abstract}
\maketitle
\section{Introduction}
Robust solid state substrates such as diamond confer an important advantage to the spin qubits hosted within: coherent operation in real-world environments at room temperature. This makes diamond quantum sensors promising probes of physical phenomena in demanding environments, such as rotation at timescales commensurate with quantum decoherence~\cite{wood_magnetic_2017, fu_sensitive_2020}. Employing color-center based spin qubits as precision quantum sensors in a range of applications is rapidly evolving, and understanding how quantum spins in solids behave when motional and rotational degrees of freedom are present is becoming of critical importance~\cite{hoang_electron_2016, delord_ramsey_2018, delord_strong_2017, delord_spin-cooling_2020, perdriat_spin-mechanics_2021}. Of the many known solid state color-centers, the nitrogen-vacancy center in diamond~\cite{doherty_nitrogen-vacancy_2013, schirhagl_nitrogen-vacancy_2014} is the most widely studied and, for quantum applications, the most eminently promising. 

While much interest is focused on the remarkable properties of the negatively-charged NV$^{-}$ center (hereafter simply NV) electron spin, the intrinsic nuclear spin hosted within the defect underpins some of the most important emerging applications. Unlike $^{13}$C nuclear spins, which are randomly distributed about the diamond lattice, the $^{14}$N or $^{15}$N nitrogen nucleus is intrinsic to every NV center, an attractive feature for scalable quantum processors and memories based on nuclear spins~\cite{fuchs_quantum_2011, zaiser_enhancing_2016} as well as multi-modal ancilla based sensing~\cite{liu_nanoscale_2019, qiu_nuclear_2021}. Featuring millisecond-long coherence times due to isolation from the environment~\cite{smeltzer_robust_2009,neumann_single-shot_2010}, the intrinsic nuclear spin is also an eminently feasible system to probe the interplay between classical rotation and quantum angular momentum, having found applications in quantum rotation sensing~\cite{maclaurin_measurable_2012,ledbetter_gyroscopes_2012, ajoy_stable_2012, jaskula_cross-sensor_2018, soshenko_nuclear_2020, jarmola_demonstration_2021} and quantum information processing~\cite{wrachtrup_processing_2006}. 

Accessing the nuclear degree of freedom and its myriad benefits in the rotating frame first requires nuclear spin hyperpolarization, which can be achieved using the electron spin. While a range of alternative methods exist~\cite{pagliero_recursive_2014,chakraborty_polarizing_2017, xu_dynamically_2019, goldman_optical_2020, gulka_room-temperature_2021, huillery_coherent_2021}, the simplest and fastest method uses a magnetic field applied precisely parallel to the N-V axis, which induces an avoided crossing in the NV excited state triplet~\cite{jacques_dynamic_2009}. At this excited-state level anti-crossing (ESLAC), $B \sim 500$\,G, the NV $m_S  = 0$ and $m_S = -1$ states approach degeneracy in the excited state. During optical illumination, coupled electron-nuclear spin states undergo spin-conserving flip-flop transitions that, with further optical cycling, result in a transfer of population to a single nuclear spin sublevel $m_I$. This process, which can be completed with a single $5$-$10\,\upmu$s laser pulse, also results in nuclear state-dependent fluorescence and enables optical quantum state readout~\cite{steiner_universal_2010, jarmola_robust_2020}.

The optical polarization procedure described above is relatively insensitive to magnetic field strength~\cite{jarmola_robust_2020}, but misalignments of less than a degree from the NV axis are sufficient to totally suppress optically-mediated nuclear spin polarization and readout~\cite{jacques_dynamic_2009}. Polarization at high magnetic fields is desirable due to the enhanced nuclear gyromagnetic ratio ($20\times$ higher at 500\,G~\cite{chen_measurement_2015} than at zero field), enabling fast quantum gate operations. As a consequence, the NV-nuclear spin system typically must be deployed with the strict condition of magnetic field alignment parallel to the NV axis. This inflexible requirement severely restricts the usefulness of nuclear spins in scenarios where the motional and rotational degrees of freedom of a quantum system are not fixed~\cite{mcguinness_quantum_2011,maclaurin_nanoscale_2013,delord_electron_2017,delord_ramsey_2018}. However, precise magnetic field alignment is only necessary for the optical polarization and readout procedure, and is not necessary to coherently \emph{control} the nuclear spin. 

In this work, we overcome the restrictive, inflexible conditions of optically-induced nuclear hyperpolarization in a truly demanding sensing environment: a frame physically rotating at 1\,kHz, faster than the quantum decoherence rate.  In our experiment, a (100)-cut diamond is mounted so that one NV orientation class makes a $54.7^\circ$ angle to the rotation axis, and a $\sim$500\,G magnetic field is aligned so that at a given initialization time, within a narrow window, the nuclear spin can be optically polarized via the ESLAC. Beyond this alignment window, the NV axis and magnetic field are free to assume any possible relative orientation dictated by the rotation of the diamond. Resonant radiofrequency fields can then be applied to drive the nuclear spins, with optical readout occurring when the NV-magnetic field alignment is reached again after one period. We are thus able to study for the first time the NV nuclear spin at high fields free from the restrictions of precise magnetic field alignment, performing first-time measurements that are virtually impossible with a stationary diamond. 

Rapid rotation in the presence of a large off-axis field represents a significant perturbation to the NV-nuclear spin system, inducing mixing between electron and nuclear spin states. The rotation is also mechanical, and thus subject to the vicissitudes of mechanically-driven processes such as period jitter and drift which invariably impact the coherence of the rotating qubits. We use feedforward control to eliminate the significant frequency shifts imposed by the off-axis magnetic field and dynamical decoupling to suppress the deleterious aspects of mechanical rotation. This quantum decoupling sequence can be made to eliminate \emph{any} detectable effects of physical rotation on-demand: preserving the nuclear spins through an entire cycle of rotation in the presence of a noisy perturbation and demonstrating the precise level of control our experiments can achieve.  

\section{Experiment}
Rapid physical rotation on the order of the NV electron spin coherence time $T_2\sim 100\,\upmu$s imposes significant but achievable demands on the stability and reproducibility of mechanical rotation. In previous work, we have demonstrated quantum control and measurement of single NV electron spins rotating at up to 200,000\,rpm ($3.33\,$kHz)~\cite{wood_quantum_2018, wood_observation_2020}. Working with ensembles of NV centers reduces some of the complexity associated with optical measurement of a rotating diamond~\cite{wood_magnetic_2017, wood_$T_2$-limited_2018}. However, magnetic field alignment during rotation for optical polarization of nuclear spins imposes challenges comparable to that of measuring single NV centers, given the strict requirements for alignment.   

A schematic of our experiment is depicted in Figure \ref{fig:fig1}(a) and more details can be found in Ref.~\footnote{See Supplemental Material.}. An electronic grade diamond (ppb N concentration, Element6) containing an ensemble of NV centers is mounted on its (100) face to the end of the shaft of a high-speed electric motor. A $480\,$G magnetic field from a permanent magnet is applied at a $54.7^\circ$ angle to the rotation axis. Spinning at $1\,$kHz, the motor controller generates a timing signal that is fed into a delay generator to trigger laser, radiofrequency and microwave pulses synchronous with the diamond rotation~\cite{wood_quantum_2018}. Adjusting the delay time effectively controls the azimuthal angle of the diamond, which is then set so that the NV and magnetic field are precisely aligned at a particular time during the rotation. A microscope objective on a three-axis piezo scanner focuses $1\,$mW of 532\,nm laser light near the exact rotation center of the diamond. The green light optically pumps the NV and its nuclear spin into the $|m_S = 0, m_I = +1\rangle$ state, and the same objective lens collects the emitted red fluorescence and directs it to an avalanche photodiode, forming a confocal microscope.
\begin{figure}
	\centering
		\includegraphics[width = \columnwidth]{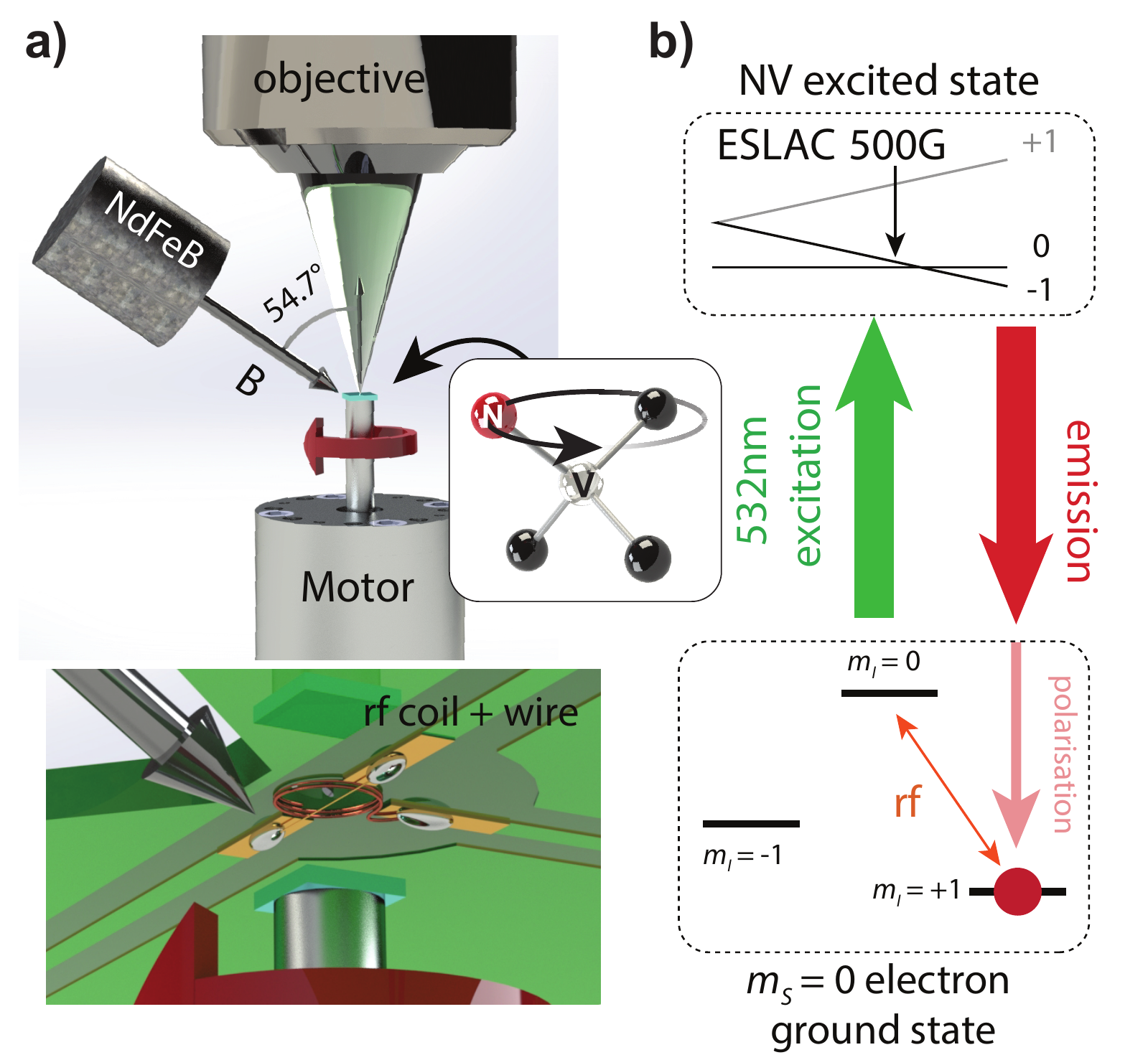}
	\caption{Rotating diamond experiment and NV energy levels (a) Schematic of the experiment: a diamond containing an ensemble of NV centers is rotated by an electric motor at $1\,$kHz. A $480\,$G magnetic field is applied so that at a particular azimuthal rotation angle, the field momentarily aligns with the NV axis, polarizing the host $^{14}$N nuclear spin. Radiofrequency (microwave) fields are generated by a coil (wire) antenna (lower panel) and are used with a 532\,nm laser to prepare, read and control the NV and its nuclear spin. (b) NV energy levels, showing nuclear spin rf transitions, excited state level anti-crossing and optical pumping process.}
	\label{fig:fig1}
\end{figure}

\section{Optical polarization}\label{sec:pol}

An NV electron spin state can be optically pumped to almost $100\,\%$ polarization within one microsecond~\cite{robledo_spin_2011, tetienne_magnetic-field-dependent_2012}. The nuclear spin polarization process is slower, typically taking up to five or more microseconds to complete~\cite{jarmola_robust_2020}. The degree of polarization also depends on the magnetic field strength~\cite{busaite_dynamic_2020}. In our experiments, the temporal dependence of optical pumping is convolved with the angular variation of the magnetic field alignment imposed by rotation, and therefore the nuclear spin polarization efficiency is also affected by rotation. To understand this effect, we study the time-dependent photoluminescence (PL) emitted from the NV during optical pumping. 

\begin{figure*}[t]
	\centering
		\includegraphics[width = \textwidth]{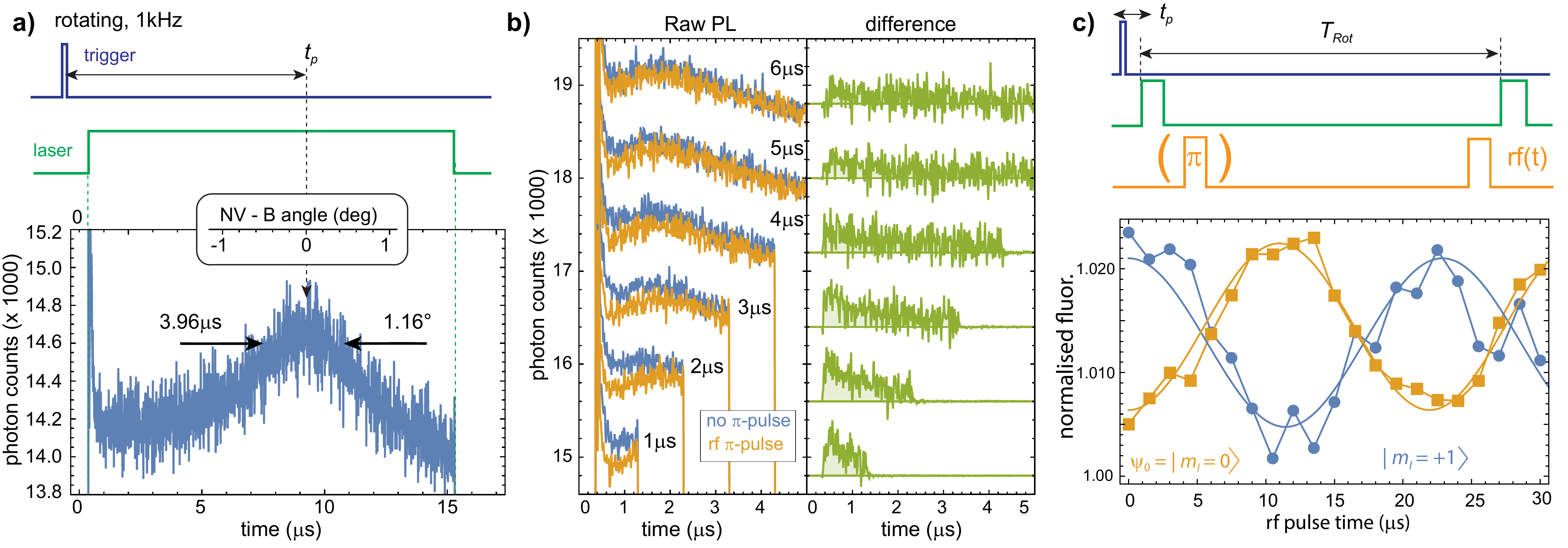}
	\caption{Quantum state measurement and control of a rotating ensemble of nuclear spins. (a) Ensemble NV PL as a function of time while rotating at 1\,kHz, with the laser switched on approximately 10$\upmu$s before the NV and magnetic field come into alignment. When this occurs, optical pumping polarises the $^{14}$NV into the $|m_S = 0, m_I = +1\rangle$ state, which exhibits brighter fluorescence. After this point, the laser depolarizes the nuclear spin, leading to a characteristic $4\,\upmu$s-wide Gaussian profile in time, equivalent to a $1.2\,^\circ$ angular window where polarization of the nuclear spin is possible. (b) State dependence of NV PL while rotating (left). Optimizing the fluorescent contrast (right) amounts to timing the laser pulse to activate on the rising edge of the Gaussian feature (approximately 2$\upmu$s from the peak) and subsequently extinguish without depolarizing the spin ensemble. We find that between 2-3\,$\upmu$s yields optimum results. (c) A variable duration pulse of resonant (5.1\,MHz) rf is applied immediately following nuclear polarization. The diamond then rotates a full period before state readout, and we observe Rabi oscillations (blue circles). We can also first apply a resonant $\pi$-pulse after polarization, which flips the nuclear spin into the dark $m_I = 0$ state before applying a variable duration rf pulse after a rotation period (orange squares).}
	\label{fig:fig2}
\end{figure*}

After first ensuring optimum optical pumping while stationary~\footnote{See Supplementary Information [URL]}, we then studied the nuclear spin polarization dynamics while the diamond was rotating. Figure \ref{fig:fig2}(a) depicts the measured photoluminescence as the NV axis approaches alignment with the magnetic field, $t_p = 8\,\upmu$s after the laser is switched on. For meaningful quantum measurement, we must detect the difference in brightness dependent on the $^{14}$N spin state. We reduced $t_p$ so that the laser switched on just before the NV axis and magnetic field are aligned, and observe a difference in fluorescence when an rf $\pi$-pulse is applied prior to readout (Fig. \ref{fig:fig2}(b)). If the laser is not promptly extinguished, the light rapidly depolarizes the $^{14}$N beyond the point where the magnetic field and the NV axis are aligned. The nuclear spin polarization is therefore limited to what is possible within the narrow window depicted in Fig. \ref{fig:fig2}(b). 

The maximum fluorescence contrast (and thus nuclear spin polarization) is markedly lower while rotating compared to stationary ($2.5\,\%$ vs. $6\,\%$). Nevertheless, we were able to perform and unambiguously detect quantum spin manipulations on the rotating ensemble. We varied the duration of the rf pulse, tracing out time-domain Rabi oscillations with the nuclear spin first initialized into the $|m_S = 0,m_I = +1\rangle$ state after optical pumping. In a further experiment, we applied a resonant $\pi$-pulse and a variable-duration rf pulse at the start and end of the cycle respectively, and thus measured Rabi oscillations from the nuclear spin initialized to the $|m_S = 0,m_I = 0\rangle$ state. The nuclear (and electron) spin adiabatically follows the quantization axis set by the NV zero-field splitting and magnetic field during the rotation. The second measurement in Fig~\ref{fig:fig2}(c) effectively confirms this adiabaticity, since the NV electron and nuclear spin state is the same at the start and end of the rotation. 

\section{Electron-nuclear spin-mixing}
\begin{figure}[t]
	\centering
		\includegraphics[width = \columnwidth]{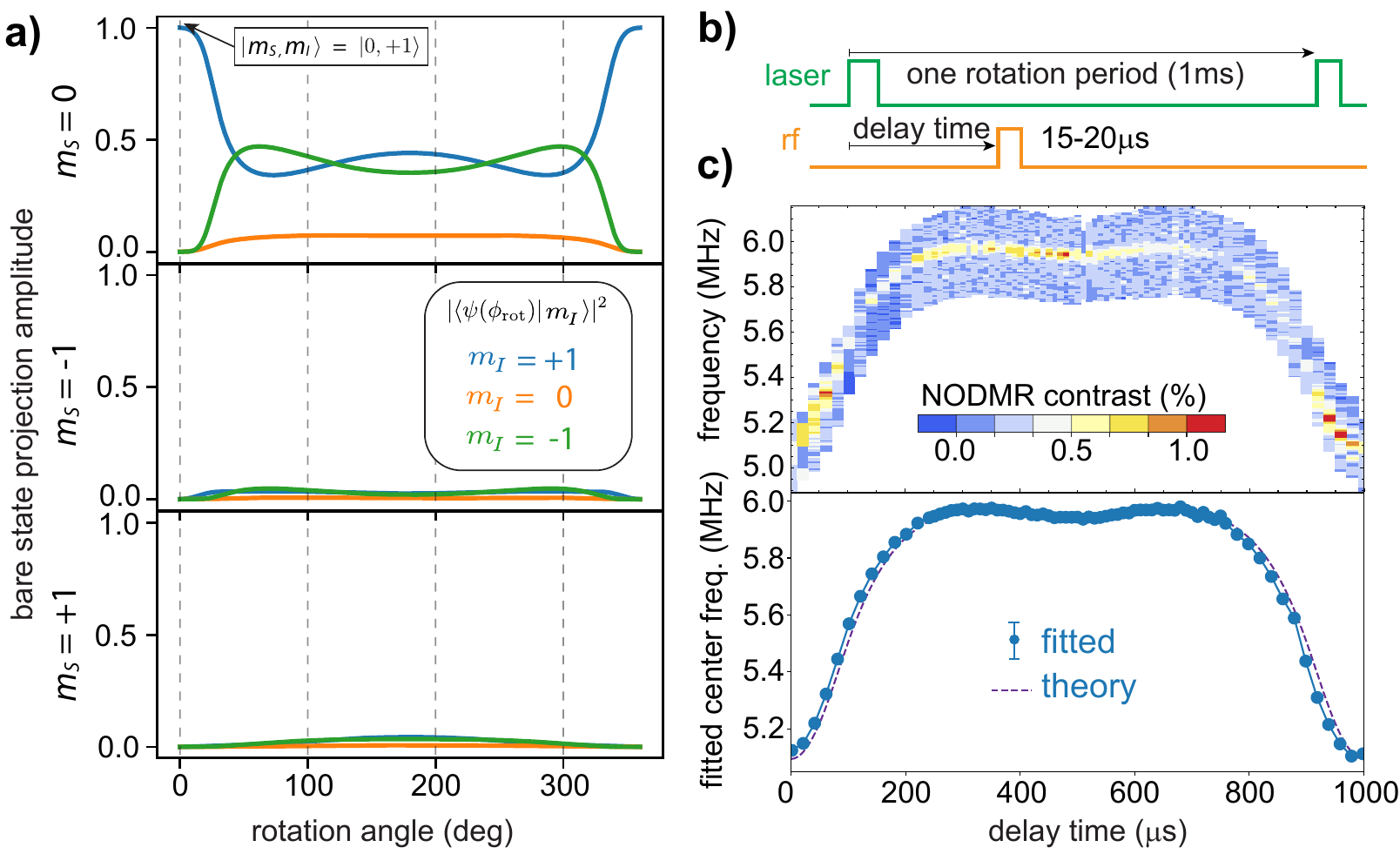}
	\caption{State mixing in the electron and nuclear spin manifolds of the NV-$^{14}$N system. (a) Initialized to the $|m_S, m_I\rangle = |0,+1\rangle$ ground state, we project the corresponding instantaneous eigenstate of $H(\phi_\text{rot})$ onto the bare eigenstates of $H$ when the field is aligned to the NV axis. (b) Experimental pulse sequence used: a laser pulse optically pumps the NV-nuclear spin, followed by an rf pulse with swept frequency is applied after a variable delay time, and after a full period the resultant spin state is read out. (c,top) NODMR contrast as a function of rf frequency and delay time. (c, bottom) Fitted center frequency extracted from Gaussian fits to the data in (b). Overlaid is the theoretical prediction derived from solving for the instantaneous energies of the system Hamiltonian. Error bars are smaller than data points, and represent standard error of fits.}        
	\label{fig:fig3}
\end{figure}

Rotation in the presence of a large field imparts a significant perturbation, effectively tilting the magnetic field from $0$ to $109^\circ$ and back, which results in strong, time-dependent mixing of the electron and nuclear spin levels. Mixing of the electron spin levels augments the nuclear spin energies, resulting in larger perturbations to the nuclear spin transition frequencies than that of the magnetic field alone. Misalignment of the magnetic field not only suppresses nuclear spin polarisation, it also quenches the fluorescence of the NV electron spin~\cite{epstein_anisotropic_2005} for much smaller fields than that used in this work. Rotation of the diamond therefore provides us with a unique means of studying both the NV and its nuclear spin in hitherto inaccessible regimes of magnetic field strength and alignment. Before further experimental investigation, we consider the Hamiltonian of the NV-$^{14}$N system and the effects of a large, time-dependent off-axis magnetic field. For the electron spin $\boldsymbol{S}$ and nuclear spin $\boldsymbol{I}$ (both spin-1), the interaction Hamiltonian is
\begin{eqnarray}
H &= D_\text{zfs} S_z^2 - \gamma_e \boldsymbol{S}\cdot\boldsymbol{B} - \gamma_n\boldsymbol{I}\cdot\boldsymbol{B} + Q I_z^2 \nonumber \\
  & + A_\parallel S_z I_z + A_\perp(S_x I_x + S_y I_y).
\label{eq:hn}
\end{eqnarray}
Here, $D_\text{zfs}/2\pi  = 2.870\,$GHz is the NV zero-field splitting, $\gamma_e/2\pi = -2.8$\,MHz\,G$^{-1}$ and $\gamma_n/2\pi = 307\,$Hz\,G$^{-1}$ the electron and nuclear gyromagnetic ratios, $Q/2\pi = -4.9457\,$\,MHz~\cite{jarmola_robust_2020} the nuclear quadrupolar splitting and $A_\parallel/2\pi = -2.162\,$MHz and $A_\perp/2\pi = -2.62\,$MHz the longitudinal and transverse coupling constants for the electron-nuclear hyperfine interaction, respectively. It is simplest to include the effects of physical rotation by considering the magnetic field as initially parallel to the NV axis and rotating about an axis $z'$. In our case $z'$ makes an angle of $54.7^\circ$ to the diamonds rotation axis due to the $\langle100\rangle$ cut. In Figure \ref{fig:fig3}(a), we project the instantaneous eigenstates of $H(t)$ on to the bare eigenvectors of $H(0)$ for the initial configuration of electron and nuclear spin state that follows immediately after optical pumping, that is, $|m_S = 0, m_I = +1\rangle$. This state adiabatically evolves into an almost equal superposition of $m_I = \pm1$ nuclear spin states, before returning to the initial state at the completion of the rotation.

To proceed further, we next consider driving the nuclear spin in the intervening regime with rf, where the NV-magnetic field angle is significant. We perform NODMR spectroscopy at different stages of the rotation by controlling the time between rf $\pi$-pulse ($t_\pi = 7\,\upmu$s when stationary) and laser readout ($1\,\upmu$s) at a rotational speed of 1\,kHz. We extract the resonance frequency from Gaussian fits to the data shown in Fig~\ref{fig:fig3}(c, top) and compare against the expected NODMR transition frequency predicted by solving for the eigenvalues of $H(t)$ for the known magnetic field strength ($480\,$G), which shows remarkably good agreement. The time offset is attributed to the extended duration of the rf pulses: while the rf Rabi $\pi$-pulse is $t_\pi \approx 7\,\upmu$s while stationary near the field alignment point, we must increase the pulse duration to $20\,\upmu$s in order to see any contrast for delay times where the magnetic field is at a significant angle. We used a coil antenna that produces rf fields parallel to the rotation axis, which should therefore result in a rotationally-symmetric rf rabi frequency. That we see significant azimuthal angular dependence of the rf coupling as the diamond rotates implies some other interaction is at work.

\section{Nuclear spin de-augmentation}\label{sec:gyraug}
To better understand why the rf coupling is reduced, we examine several key aspects of how the nuclear spin, augmented by the NV electron spin, couples to the rf field. At the ESLAC and when $B$ is parallel to the NV axis, the observed rf Rabi frequencies are $\alpha_0\sim20$ times what would be predicted based purely on the $^{14}N$ gyromagnetic ratio, $\gamma_{14N}/2\pi = 307.7\,$Hz/G, due to the presence of the NV electron spin ~\cite{chen_measurement_2015, sangtawesin_hyperfine-enhanced_2016}. The electronic augmentation stems from the transverse hyperfine coupling interaction, which is quantified by the constant $A_\perp$. In the presence of an off-axis magnetic field, the effective hyperfine coupling has different components parallel and perpendicular to the new quantization axis $z'$, which is no longer parallel to the NV axis. The Hamiltonian $H$ in Eq. \ref{eq:hn} is diagonalized by the transformation $P^{-1}H P = E$, with $P$ the matrix of eigenvectors and $E$ a diagonal matrix of energy eigenvalues. To reveal the augmentation of the nuclear spin gyromagnetic ratio, we transform the rf coupling Hamiltonian $H_\text{rf} = B_\text{rf}(t)\left(\gamma_e S_x + \gamma_N I_x\right)$ into the basis where $H$ is diagonal, $H'_\text{rf} = P^{-1} H_\text{rf}P $. Eliminating the time dependence of $H_\text{rf}$ with the rotating-wave approximation and computing the matrix element corresponding to the nuclear spin transition of interest, in this case, $|m_S = 0, m_I = +1\rangle\rightarrow|0,0\rangle$, we have
\begin{equation}
\alpha_0 = \frac{1}{\gamma_N B_\text{rf,0}} \langle 0,0|H'_\text{rf}|0,+1\rangle
\end{equation}
for $B||z$ and $B_\text{rf,0}$ the amplitude of the oscillating field. When the diamond is rotated, the increasingly off-axis magnetic field tilts the quantization axis further from the NV axis. Since the nuclear spin eigenenergies never approach degeneracy (and the rotation rate $f = 1/T$ is slow and thus adiabatic), we can denote the instantaneous eigenstates as $|\eta(t)\rangle$ and $|\zeta(t)\rangle$, with $|\eta(0)\rangle = |\eta(T)\rangle = |0,+1\rangle$ and $|\zeta(0)\rangle = |\zeta(T)\rangle = |0,0\rangle$. We then extract the effective augmentation factor 
\begin{equation}\label{eq:gyraug}
\alpha'_0(t) = 1/\gamma_N B_\text{rf,0} \langle \zeta(t)|H'_\text{rf} (t)|\eta(t)\rangle.
\end{equation}

In our previous experiment, the rf field was roughly parallel to the rotation axis. A purely geometric argument, whereby $\Omega(t)\sim |\uvect{n}(t)\times\boldsymbol{B}_\text{rf}|$, would have the Rabi frequency remain constant across the rotation. The tilted quantization axis due to the magnetic field results in a significant reduction of the rf Rabi frequency through a rotation cycle. Our experiment gives us a unique opportunity to detect this effect, given that the angle between the NV and magnetic field is free to assume any angle after (and before) optical polarization (readout). In the next section, we compare experimental measurements of this effect with theory. 

\section{Feed-forward quantum control}
To further characterize the time/angle dependence of gyromagnetic augmentation, we would ideally measure the rf Rabi frequency across the full rotation in an experiment. However, the large, rapid changes in the transition energy (Fig. \ref{fig:fig3}(c)), far exceeding our maximum achievable Rabi frequency ($70\,$kHz), preclude straightforward investigation. We tackle this problem by applying feed-forward quantum control. With the data in Fig. \ref{fig:fig3}(c) as a guide, we can synthesize a sine wave signal, frequency modulated in a manner set by the angular variation of the two-level splitting using an arbitrary waveform generator (AWG). This feed-forward ensures the rf frequency is always close to resonance with the nuclear spin throughout the rotation cycle, as shown in Fig. \ref{fig:fig6}(a), eliminating to a significant extent the strong modulation of the transition energy imparted by the off-axis magnetic field. 

We determine the feedforward waveform by first measuring the frequency $f_0$ of the nuclear spin transition (while stationary) to a precision of better than 100\,Hz with optimized magnetic field alignment. From this, we then compute the FM profile by diagonalizing Eq. \ref{eq:hn} for the equivalent magnetic field $B_0$ at each azimuthal angle corresponding to the experiment time, assuming that at the peak alignment point $\theta_B\approx0$ and $\theta_{NV} = 54.7^\circ$. Using just this simple assumption, we can obtain high-fidelity rf pulses at any point of the rotation. 

\begin{figure}
	\centering
		\includegraphics[width = \columnwidth]{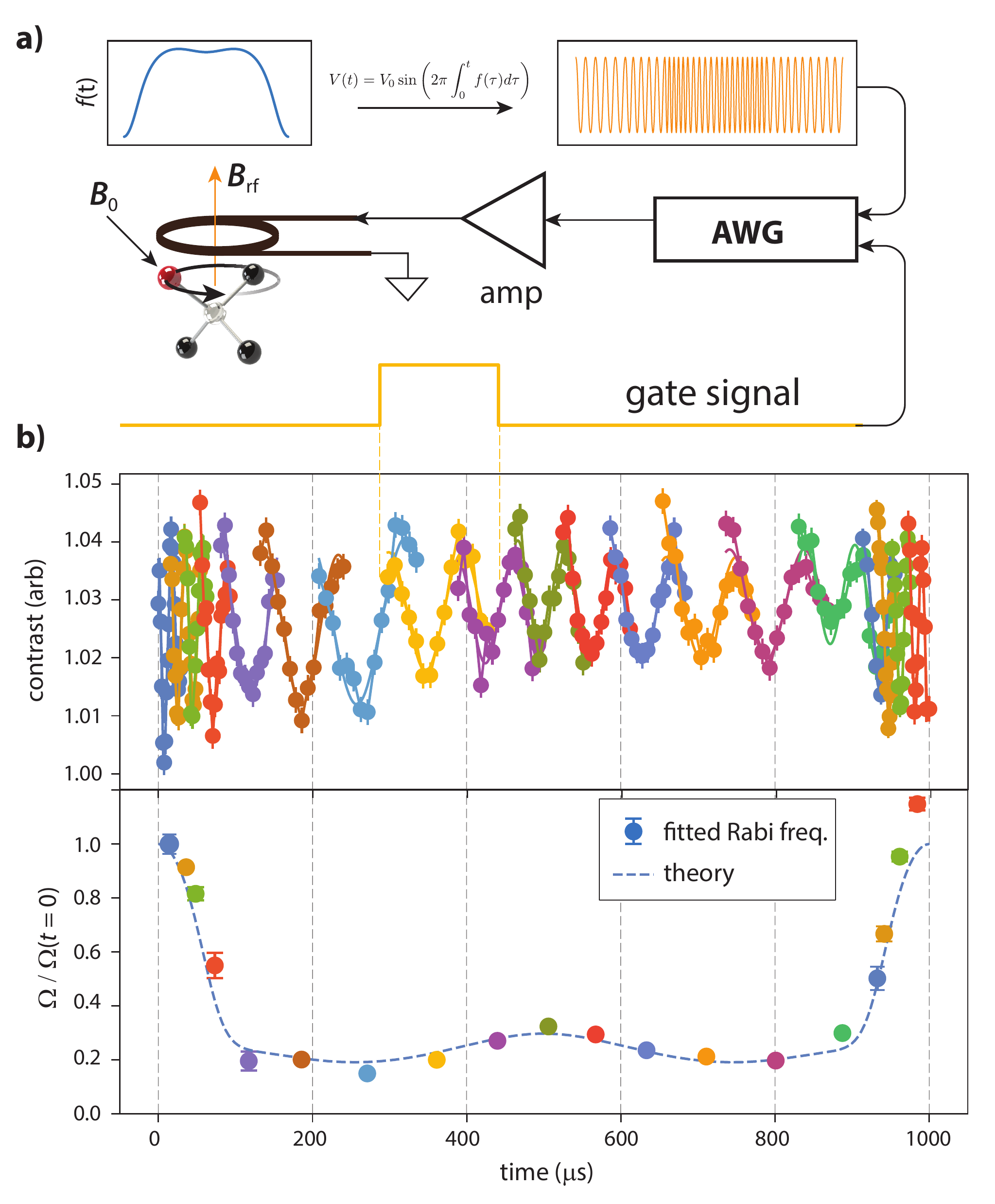}
	\caption{Feed-forward quantum control of nuclear spins in a rapidly rotating diamond. (a) Using the observed variation of the $|0,+1\rangle\rightarrow|0,0\rangle$ transition frequency with rotation angle, we generate a frequency-modulated rf signal using an AWG that ensures the rf is always resonant with the nuclear spin transition. The rf is digitally gated synchronous with the rotation after a delay time $t_d$, and Rabi oscillations are traced out by varying the gate time. (b) We are able to drive near-resonant Rabi oscillations at any point in the rotation which markedly slow down due to the time-varying electron-nuclear spin augmentation. Error bars derive from photon counting statistics. (c) The relative variation of the Rabi frequency matches well with a fully-constrained model of the effective nuclear gyromagnetic ratio, based on Eq. \ref{eq:gyraug}. Error bars represent uncertainty in fitted Rabi frequencies.}        
	\label{fig:fig4}
\end{figure} 

We then initiate Rabi oscillations at a delay time $t_D$ with a variable-duration gated rf pulse from the AWG. The results are shown in Fig. \ref{fig:fig4}(b), and in Fig. \ref{fig:fig4}(c) we plot the extracted Rabi frequency. Our experimental results are well explained by the model and the arguments based on the loss of gyromagnetic augmentation induced by an off-axis magnetic field, Eq. \ref{eq:gyraug}. Despite the reduced rf coupling, we are able to control the quantum state of the nuclear spin at essentially any point throughout the rotation, which enables us to then deploy interferometric measurement sequences to quantify the coherence of the rotating nuclear spin.

\section{Coherence of rotating nuclear spins}
The results of the previous section allow us to map out the angular variation of the rf Rabi frequency and adjust the pulse durations in a multiple-pulse sequence such as Ramsey interferometry, or spin-echo. As the control signal is generated by an arbitrary waveform generator, we have the freedom of absolute phase control of the rf at any point of the rotation. To measure the ensemble dephasing time $T_2^\ast$ of the nuclear spins, we use a Ramsey pulse sequence, $\pi/2 - \tau - \pi/2$ and vary the phase of the final rf $\pi/2$-pulse to trace out fringes. This way, phase shifts induced by an imperfect feedforward profile do not affect the perceived fringe visibility. The spin coherence time $T_2$ is measured using a $\pi/2 - \tau / 2 -\pi - \tau/2 - \pi/2$ spin-echo sequence, where again the phase of the final rf pulse is varied to measure the fringe amplitude. 

\begin{figure*}
	\centering
		\includegraphics[width = \textwidth]{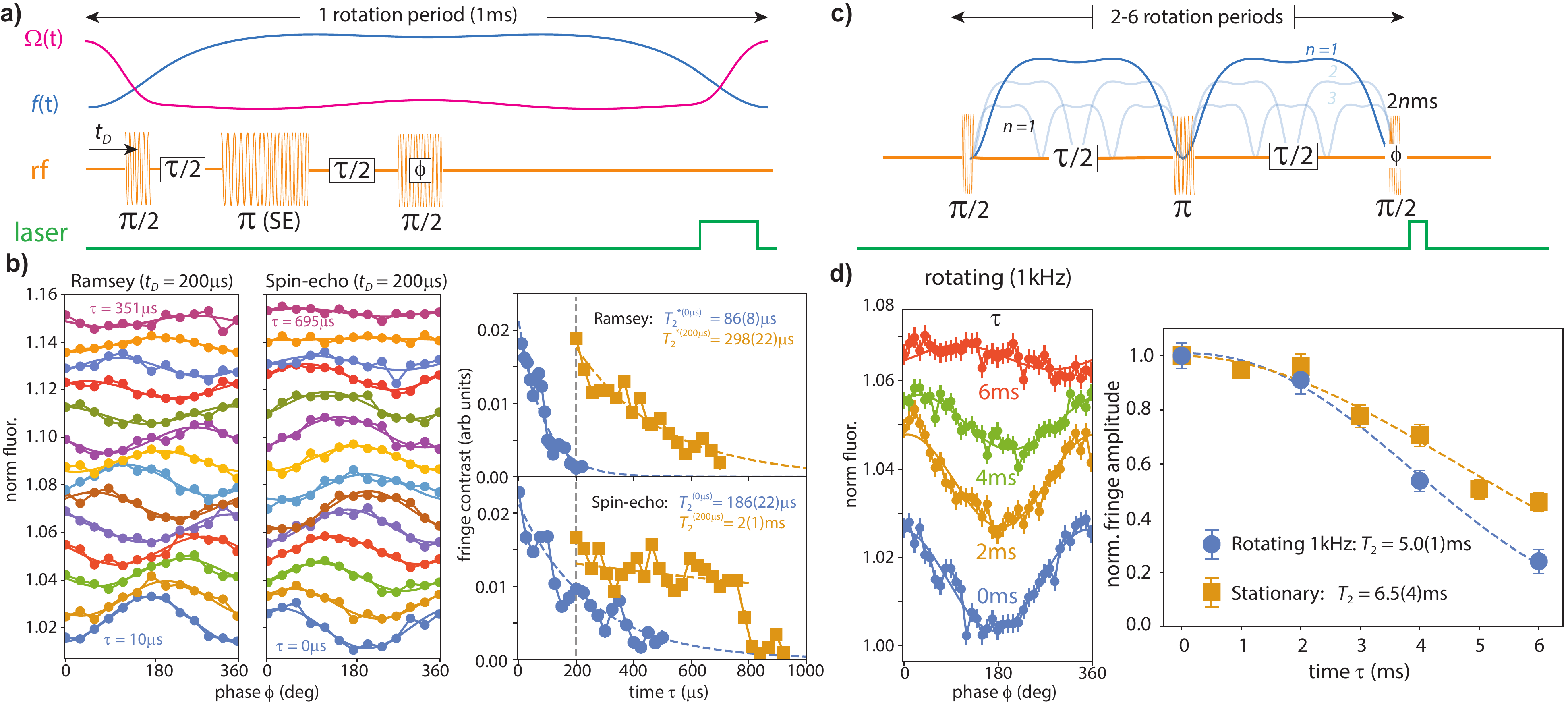}
	\caption{Quantum coherence measurement of rapidly rotating nuclear spins. a) Pulse sequence: using feedforward control of the rf frequency, and adjusting rf pulse lengths to execute $\pi/2$ or $\pi$-pulses, we perform Ramsey or spin-echo interferometry on the rotating nuclear spin ensemble. The phase of the final $\pi/2$-pulse is varied to trace out fringes (b). The spin coherence times while rotating are then derived by fitting an exponential decay to the fringe amplitudes as a function of interrogation time $\tau$. In (c), we do the same spin-echo measurement but restrict $\tau$ to twice-integer multiples of the rotation period (1\,ms). Phase differences accumulated in subsequent periods cancel, and we recover the spin-echo signal (d). The observed spin coherence time $T_2 = 5.0(1)$\,ms is close to the measured stationary value of $6.5(4)$\,ms. Error bars in signal data derive from photon counting statistics, and error bars on fringe amplitudes are standard error in the fitted sinusoidal amplitude.}        
	\label{fig:fig5}
\end{figure*}

An additional control parameter is the time the pulse sequence starts at, $t_D$, which allows us to select the state superpositions used in the interferometry sequence. Figure \ref{fig:fig5}(a,b) shows the pulse sequence, Ramsey and spin-echo data with the extracted fringe phase and amplitude for $t_D = 0$ and $t_D = 200\,\upmu$s. We observe that when the interferometric interrogation region includes the rapid change in state populations around the time where the magnetic field and NV axis come into alignment (0 - 200\,$\upmu$s and 800 - 1000\,$\upmu$s), the coherences drop sharply, $T_2^\ast = 86(9)\,\mu$s and $T_2 = 186(22)\,\mu$s. When the Ramsey or spin-echo sequence excludes these regions, the coherence times are much longer: $T_2^\ast = 299(22)\,\mu$s and $T_2 = 2(1)$ms. A particularly stark example of this difference being for spin-echo starting at 200\,$\upmu$s: around $800\,\upmu$s in Fig. \ref{fig:fig5}, the state populations begin to change rapidly and the spin-echo signal quickly damps~\footnote{Due to the long rf pulse durations ($t_\pi \approx 50\,\upmu$s), $\tau = 800\,\upmu$s corresponds to the final rf pulse occurring almost $900\,\upmu$s into the rotation.}.

While excluding the rapidly varying regions improves the coherence times, our observed rotating coherence times are still well below that observed when the diamond is held stationary ($T_2^\ast = 1.5\,$ms, $T_2 =6.5(4)$ms). In Fig. \ref{fig:fig5}(c), we increase the spin-echo time in two-period intervals ($2\,$ms), with the rf pulses applied only when the NV axis and magnetic field are aligned ($t_D = 0$). Here, we see the contrast persists to times commensurate with what is observed while stationary, $T_2 = 5.0(1)$ms, demonstrating coherent quantum measurement of nuclear spins physically rotating at 1\,kHz.  

With these results, we can begin to piece together the effect of rotation on the coherence of the NV nuclear spin. The data in Fig. \ref{fig:fig5}(c,d) are compelling evidence that the intrinsic coherence of the nuclear spin is unaffected by adiabatic physical rotation in the presence of a large off-axis magnetic field. However, the rotation is not necessarily perfect, and spin-echo contrast observed at 2$n$ period intervals only means that the period-to-period variability of the rotation speed is small, since whatever (noisy) phase accumulated in the first period is canceled in the second period that immediately follows it. Moreover, if the angular velocity within a period is not constant, such that the NV axis traces out an ellipse or more complicated parametric shape each period, phase effects from this effect will aslo be eliminated with spin-echo at two-period intervals. Loss of coherence at non-periodic intervals does not automatically imply that our feedforward frequency control is imperfect either, since for coherence to vanish, noise and irreproducibility must be present. 

The most obvious source of noise is the stability of the diamond rotation. We performed an independent measurement of the motor stability using laser detection of a fiducial \footnote{See Supplementary Information} and found that the period exhibited a typical standard deviation of 323\,ns. Even such small period jitter has a disproportionately severe effect on the relative phase between the local oscillator (the AWG waveform) and the NV due to the large perturbation induced by the magnetic field, in particular the rapid vartiation in transition frequency close to the magnetic field alignment point. A linear frequency increase results in a quadratic increase in integrated phase, and while this is partially canceled by the feedforward waveform, small differences in the nuclear spin phase due to period jitter are nevertheless amplified quadratically and result in a loss of phase coherence between the start and end of the frequency sweep. After $t_D = 200\,\upmu$s, the phase increases approximately linearly (the transition frequency is almost constant in time), and thus coherence loss from period jitter is suppressed. When the nonlinear phase accumulation begins towards the end of the period (\emph{i.e.} total sequence times $>800\,\upmu$s), we again see a sharp drop in coherence.        

\section{Continuous dynamical decoupling of an ensemble of strongly-perturbed nuclear spins}
The detrimental effect of motor period jitter can be eliminated by working in a frame where phase accumulation is determined entirely by the phase coherence of the rf field, dynamically decoupling the nuclear spin from its environment. Since the rf Rabi frequency varies across the rotation due to gyromagnetic augmentation (Sec. \ref{sec:gyraug}), multi-pulse dynamical decoupling is nontrivial to implement without risk of introducing significant pulse fidelity issues. We instead opt to employ continuous dynamical decoupling, or spin-locking, in conjunction with quantum feedforward control. A pure $|0,+1\rangle$ is prepared at the NV axis-magnetic field alignment point, an rf $\pi/2$-pulse is then applied followed immediately by a variable duration spin-locking rf pulse $90^\circ$ out of phase to the first rf pulse. After a variable spin-locking time, a final $\pi/2$-pulse (calibrated to the requisite length based on gyromagnetic augmentation) is applied, and its phase varied to reveal fringes. 

\begin{figure}
	\centering
		\includegraphics[width = \columnwidth]{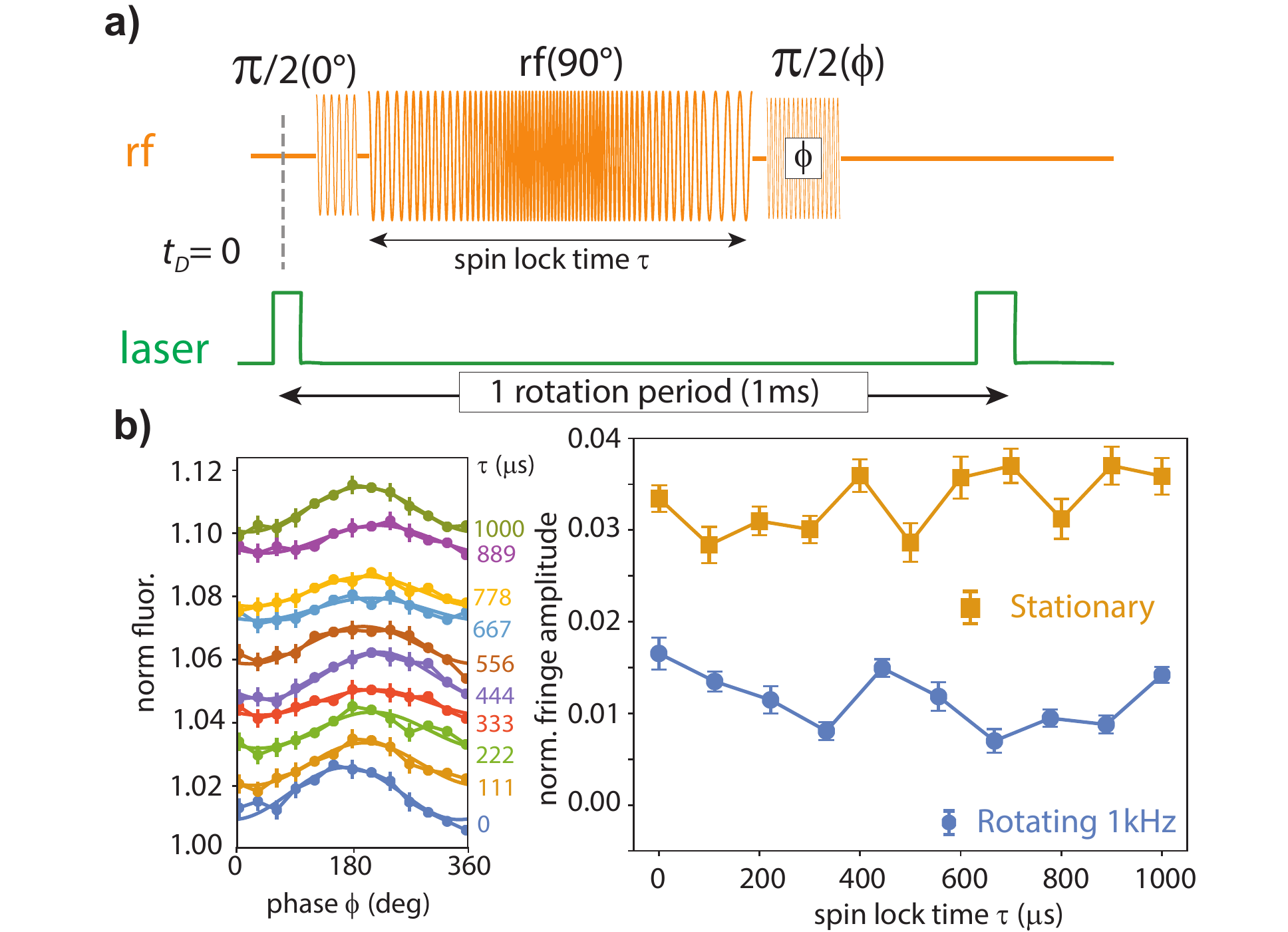}
	\caption{Spin-lock dynamical decoupling of rotating nuclear spins. (a) AWG pulse sequence. While generating a feedforward rf signal synchronous with the diamond rotation, the output is gated into three pulses, a $\pi/2$ pulse (zero phase), a variable duration pulse $90^\circ$ out of phase to the initiation pulse and a final $\pi/2$ pulse with variable phase and a duration set by the instantaneous Rabi frequency (Fig.\ref{fig:fig4}). Laser pulses before (after) prepare (read) the NV spin state. (b) Sample spin-lock signal fringes for rotation at $1\,$kHz. Error bars derive from photon counting statistics. (c) Fitted fringe amplitude as a function of spin-locking time for rotating and stationary diamond. Error bars are standard error in fitted fringe amplitudes.}        
	\label{fig:fig6}
\end{figure}

The results of spin-lock control for rotation at $1\,$kHz are depicted in Figure \ref{fig:fig6}, and compared against stationary results. The lower amplitude of the rotating data is due to the deficiencies of state polarization discussed in Sec. \ref{sec:pol}. The slight dips in contrast coincide with lower Rabi frequency regions (Fig. \ref{fig:fig4}(c)), we thus attribute the contrast dip to imperfect readout ($\pi/2$-pulse) calibration. We also observe a small but measurable phase change across the rotation, which we attribute to a time-dependent detuning introduced through imperfect calibration of the feedforward profile. Otherwise, the stationary and rotating spin-lock measurements are in remarkable agreement, with the spin coherence and phase before and after a rotation virtually the same. The data in Fig. \ref{fig:fig6} demonstrate coherent decoupling of the nuclear spin while adiabatically rotating in the presence of a significant, noisy, time-dependent perturbation; that is, the off-axis magnetic field. The combination of feedforward quantum control and spin-locking allow us to eliminate the significant effects of the off-axis magnetic field, jitter of the rotation period, and effectively even the rotation itself, since the nuclear spin accumulates essentially no phase during the rotation while under spin-lock control.  
	
\section{Discussion}
This work represents the first study of the effects of rapid physical rotation and significant magnetic field angles on the quantum properties of the $^{14}$N nuclear spin in diamond. Magnetic fields used in this work allow rapid polarization of the nuclear spin via the NV excited state level anticrossing, and an interesting new regime to probe how off-axis magnetic fields affect the nuclear spin where mixing with the electron spin is significant. Optical polarization and readout of nuclear spins can still be achieved while rotating at periods much shorter than the ensemble dephasing time of the nuclear spin, and standard quantum state manipulations ($\pi/2$, $\pi$-pulses) can be carried out with high fidelity. The large perturbation imposed by the off-axis magnetic field leads to considerable state mixing within each electron spin manifold of the NV, resulting in a $20\,\%$ modulation of the nuclear spin transition frequency as the diamond is rotated. We counter this effect with feedforward quantum control, modulating the rf waveform that drives the nuclear spins with the same characteristic function imposed by the magnetic field. 

With feedforward control, the rf and two-level system remain in resonance despite the significant perturbation of the off-axis magnetic field, and for the first time we can observe how the NV nuclear spin behaves in such circumstances. A key observation we make is that the electronic augmentation of the nuclear gyromagnetic ratio depends sensitively on the alignment of the magnetic field, and as a result the nuclear spin Rabi frequency drops by almost an order of magnitude throughout the rotation. Nevertheless, we retain coherent control over the nuclear spin with appropriately tailored FM quantum control pulses that still execute high-fidelity gate operations. Using feedforward control and the standard quantum sensing protocols of Ramsey and spin-echo interferometry, we observe the effects of noisy, imperfect rotation on the coherence of the nuclear spin, and with spin-locking, we eliminate these effects as well, demonstrating almost complete decoupling of the nuclear spins from the significant perturbations imposed by rotation.  

We attribute the reduced coherence times we see when performing interferometry within a single period as evidence of imperfect rotation, a conclusion reinforced by the observation that coherence remains when the interrogation time is set to an even multiple of rotation periods. Imperfect rotation couples to the coherence of the qubits magnetically, via the perturbed transition frequency induced by rotation in the presence of an off axis bias magnetic field. While feedforward eliminates a significant fraction of the relative phase between the rf and the qubits, noise in the form of period jitter of the motor makes the relative phase vary with each iteration of the experiment, an effect accentuated when the transition frequency varies sharply (\emph{i.e.} at the beginning and end of the rotation). This effect is the same for spin-echo and Ramsey, and in spin-echo the sensitivity can in fact be higher depending on the difference between feedforward signal and qubit transition frequency.

One avenue of improvement is an optimization procedure for the feedforward waveform. To accommodate the possibility of non-circular rotation, we require an `on-the-fly' means of optimizing the feedforward profile using quantum measurement of the nuclear spins while rotating. However, this is a non-trivial undertaking, given the subtle, more interesting effects of rotation such a procedure may erase. There are two non-trivial quantum phases that the qubit accumulates while rotating that are related to the geometric Berry phase~\cite{berry_quantal_1984, aharonov_phase_1987}. The first of these relates to the path the rf field traces out in real space, and its subsequent appearance in projective Hilbert space (the Bloch sphere), which we reported on previously in~\cite{wood_observation_2020}. The second is the `conventional', state-dependent Berry phase equivalent to the solid angle swept out by the quantization axis of the system (tilted from the N-V axis by the magnetic field) as the diamond is rotated. In a rotating quantum interferometry sequence, we detect an indistinguishable combination of these `interesting' physical effects and more mundane effects such as an offset or skew in the feedforward profile compared to the transition frequency. Before a rigorous, definitive quantum measurement of rotationally-induced phases can be undertaken, further work is needed to better understand how the desired phase and systematic effects can be distinguished.

The unique capabilities developed in this work lay the foundations for several interesting avenues of future inquiry. The strong off-axis field regime of the NV is also generally inaccessible due to quenched optical contrast~\cite{epstein_anisotropic_2005}. While control of the NV electron spin with microwaves is not considered here, it could bring new insights to our understanding of the NV coupling to strain and electric fields in the high transverse field regime, and considerable richness to future work employing nuclear spins in the rotating frame. Another intriguing possibility is employing other nuclear spin species, such as $^{13}$C, which couple to each other via the dipolar interaction in natural abundance samples. An interesting interplay is then expected between NV-mediated nuclear hyperpolarization and spin diffusion~\cite{wunderlich_optically_2017, pagliero_multispin-assisted_2018} and rotational averaging of homonuclear couplings, since the experimental configuration in our work emulates that of NMR magic-angle spinning~\cite{andrew_nuclear_1958, wood_anisotropic_2021}. 

Devoid of the nuclear quadrupole interaction exhibited by the spin-1 $^{14}$N nucleus, the ground state splitting of the $^{15}$N is only $\sim200$\,kHz, allowing exploration of the regime where rotation speed, Rabi frequency and Larmor frequency are all comparable. Additionally, the nuclear spin state mixing observed in this work would be radically different in the case of $^{15}$N, with the nuclear spin more inclined to follow the external magnetic field due to the absence of the intrinsic quadrupole axis. Finally, extension to the ground-state level anticrossing (GSLAC) and the consequent effects on gyromagnetic augmentation~\cite{sangtawesin_fast_2014, sangtawesin_hyperfine-enhanced_2016, chen_measurement_2015} would be possible with a stronger magnetic field than used in our experiments.   

\section{Conclusions}  
In this work, we have demonstrated quantum measurement and control of an ensemble of nuclear spins rotating in the presence of a large, off-axis magnetic field. We have shown, for the first time, how quantum feedforward of the rf drive can be used to maintain coherent control of the nuclear spins even when the magnetic field is orthogonal to the NV axis. Furthermore, we reveal how imperfect rotation couples to the coherence of the nuclear spins, and using continuous dynamical decoupling, preserve the coherence of the spin ensemble over a full rotation period. Our work highlights the unique flexibility the NV center in diamond brings to quantum sensing, allowing access to a long-lived nuclear quantum memory embedded within a robust substrate that can be rapidly rotated at quantum coherent timescales. We also harness the power of dynamical decoupling to suppress all traces of the large magnetic perturbation and noisy, imperfect rotation, effectively giving us the ability to induce physical rotation on-demand. This work unlocks a previously frozen degree of freedom for the NV and its host nuclear spin, opening a wide range of possibilities for practical quantum sensing~\cite{soshenko_nuclear_2020, jarmola_demonstration_2021} and fundamental quantum science with rotating systems~\cite{delord_electron_2017, delord_ramsey_2018, delord_spin-cooling_2020, wood_observation_2020, pellet-mary_magnetic-torque_2021, perdriat_angle_2021}.    
   
\section*{Acknowledgements}
We thank L. C. L. Hollenberg, J.-P. Tetienne and L. D. Turner for insightful discussions. This work was supported by the Australian Research Council (DP190100949). A. A. W. acknowledges valuable support from the UoM Albert Shimmins Research Continuity Scheme.

\end{document}